\documentclass[FBSedit,FBSmath,ecsub]{FBSsuppl}
\usepackage{amsfonts}
\usepackage{amssymb}
\usepackage{epsfig} 

\title{Solutions of Faddeev-Yakubovsky equations in configuration space
for the 4N scattering states}
\author{Jaume Carbonell}
\institute{Institut des Sciences Nucl\'eaires,
           53 Av. des Martyrs, 38026 Grenoble, France}

\begin{document}

\maketitle
\begin{abstract}
The Faddeev-Yakubovsky equations in configuration space have been solved 
for the four nucleon system with special interest in their scattering states.
We present results concerning the structure of the first $^4$He,
scattering states in different (T,S) channels
including the first inelatic threshold N+3N$\rightarrow$NN+NN
and the n-t cross section using realistic potentials. 
\end{abstract}

\section{Introduction}

\begin{figure}[hbtp]
\begin{center}\epsfig{file=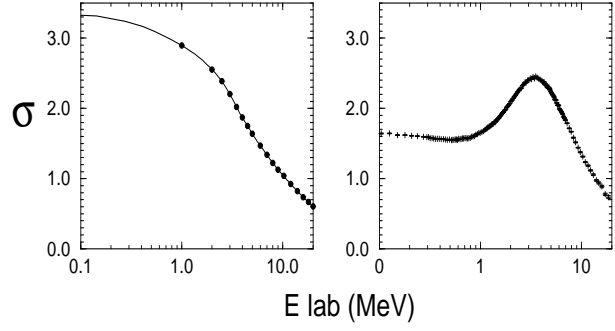,width=8cm,height=4.5cm}\end{center} 
\vspace{-0.8cm}
\caption[]{Comparison between the nd and nt elastic cross sections (in barns)}\label{nd_nt}
\end{figure}

The theoretical description of the A=4 continuum
constitutes, in addition to its technical difficulty, a serious challenge for the NN interaction models.
The remarkable success encountered in the description of the 3N system and in the
4N bound state \cite{GWKHG_AIP_95} is largely based on the three nucleon interaction (TNI).
The two body dynamics alone looses $\approx10\%$  in the 3N binding energy 
and $\approx15\%$  in the 4N.
TNI compensate this defect but, a part from an energy scaling, 
are found to have small influence in the low energy scattering states \cite{WGHGK_PRL_98}.
Contrary to the A=3 case, these states present several  nearthreshold structures,
which can be hardly
sensible to the TNI and should be reproduced with the two-body forces only.
A first example is provided by the n+t cross section which
manifests a much more vivid behaviour than the flat n-d one, 
with a resonant structure at $T_{lab}\approx3$ MeV (figure \ref{nd_nt}).
A second interesting accident is the existence of quasidegenerate n-$^3$He and p-$^3$H thresholds
with, in their middle, the first $^4$He excitation and all that in an energy gap of $0.76$ MeV.  
The position of such a state, which dominates the inter-threshold
dynamics, has to be ensured with an accuracy better than 100 KeV in order to reproduce
the experimental cross sections,
what is a strong requirement for an excited state.

\section{Formalism}\label{Sec_Formalism}

In the case of four identical particles interacting via a two-body  potential V,
the Faddeev-Yakubovsky (FY) equations result into two integrodifferential equations,
coupling two amplitudes denoted K and H:
\begin{eqnarray}
(E-H_0-V)K &=&V\left[(P_{23}+P_{13})\;(\varepsilon+P_{34})\;K  +\varepsilon(P_{23}+P_{13})\; H\right]\cr
(E-H_0-V)H &=&V\left[(P_{13}P_{24}+P_{14}P_{23}) \;K +
P_{13}P_{24}\; H\right]  \label{FYE}
\end{eqnarray}
in which $P_{ij}$ are the permutation operators and $\varepsilon=\pm1$ 
depending on we deal with bosons or fermions.
The total wavefunction is reconstructed according to:
\begin{eqnarray}
     \Psi  &=&  \Psi_{1+3} + \Psi_{2+2} \cr  
\Psi_{1+3} &=&  \left[ 1+ \varepsilon(P_{13}+P_{23}       ) \right]\;
                \left[ 1+ \varepsilon(P_{14}+P_{24}+P_{34}) \right]K\label{psi13}\\
\Psi_{2+2} &=&
\left[ 1+\varepsilon( P_{13}+ P_{23}+ P_{14}+ P_{24} ) + P_{13}P_{24}\right]\;H  \label{psi22}
\end{eqnarray}
Each amplitude F=K,H considered as a function of its own set of Jacobi coordinates,
is expanded in the corresponding angular variables according to
\begin{equation}\label{KPW}
\langle\vec{x}\vec{y}\vec{z}|F\rangle=
\sum_{\alpha} \; {F_{\alpha}(xyz)\over xyz} \;Y_{\alpha} (\hat{x},\hat{y},\hat{z})  .
\end{equation}
$F_{\alpha}$ are the unknowns and $Y_{\alpha}$ are 
tripolar harmonics containing spin, isospin and angular momentum variables.
Label $\alpha$ holds for the set of quantum numbers defined in a given coupling scheme and includes
the type of amplitude K or H.

Boundary conditions for scattering states are of Dirichlet-type. In the 1+3 elastic case e.g. 
they are implemented by imposing at large enough value of $z$  
\begin{eqnarray*}
K(x,y,z) &=&  t(x,y) \\
H(x,y,z) &=&  0
\end{eqnarray*}
$t(x,y)$ being the triton Faddeev amplitudes previously determined.
FY equations ensure a solution which, for n+t S-waves, behaves asymptotically like
\begin{eqnarray*}
K(x,y,z) &\sim&  t(x,y)\sin{(qz+\delta)}
\end{eqnarray*}
where $\delta$ is the phaseshift and $q$
is related to the center of mass kinetic energy $T_{\mathrm{cm}}$ and to the physical momentum $k$ by
\(T_{\mathrm{cm}}={\hbar^2\over m}q^2={2\over3}{\hbar^2\over m}k^2\).
A more detailed explanation of the formalism and numerical methods used
can be found in \cite{CC_98}.

\section{Results}

\begin{figure}[hbtp]
\begin{center}\epsfig{file=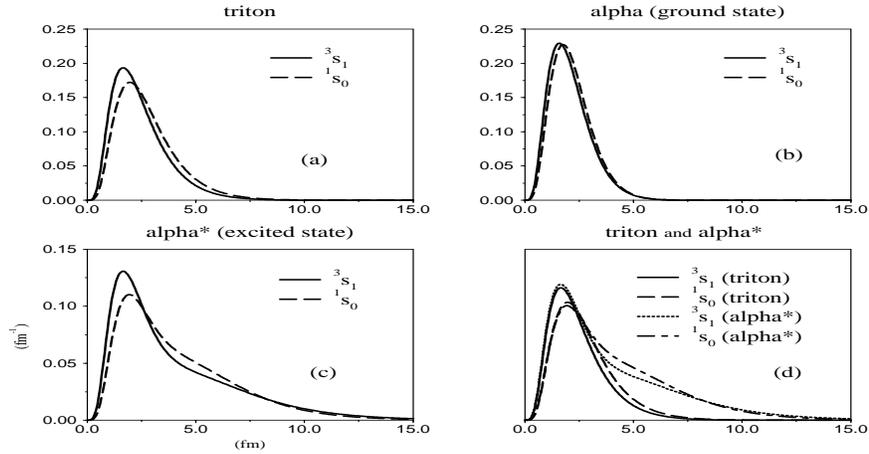,width=11.5cm,height=6cm}\end{center} 
\vspace{-0.50cm}
\caption{Two-body correlations for (a) triton, (b) $^4$He ground and (c) first excited states $^4$He$^*$.
Solid (dashed) line denotes the triplet (singlet) contributions. In (d), the results of $^4$He$^*$
are compared to the triton correlation function suitably scaled.}\label{figcorrel}
\end{figure}

We have been interested in the first $0^+$ excitation of $^4$He, 
experimentally manifested as a $\Gamma$=0.5 MeV resonant state, 
0.4 MeV above the pt threshold \cite{TWH_92}.
Using MT I-III potential \cite{MT_69} we got \cite{CC_98} the binding energies (B) and r.m.s. radius (R)
displayed in table \ref{tab_He_MT},
\begin{table}[htbp]
\begin{minipage}[h]{46mm}
\caption{Binding energy (MeV) and r.m.s. radius (fm)}\label{tab_He_MT}
\begin{tabular}{llll}  
  &  $^4$He  & $^4$He$^*$ & $^3$H  \\  \hline
B &  30.30   &  8.79	  &  8.53  \\
R &   1.44   &  4.95	  &  1.72  
\end{tabular}
\end{minipage}\hspace{0.3cm}
\begin{minipage}[h]{70mm}
\caption[]{Low energy N+3N parameters (fm)}\label{tab_lep}
\begin{tabular}{ccrrrr}
S&T&   $a$    &  $r_0$      & $v_0$          & $k_0$  \\ \hline
0&1& 14.75    &  6.75       & 0.462          & -	            	 \\
1&1&  3.25    &  1.82       & 0.231          & -	            \\
0&0&  4.13    &  2.01       & 0.308          & 0.505	            	  \\
1&0&  3.73    &  1.87       & $\simeq$ 0     & -                  	  
\end{tabular}
\end{minipage}
\end{table}
what means a bound state with B$_{^4He^*}$=0.26 MeV below the N+NNN threshold.
Using AV14 \cite{AV14_84} and Nim II \cite{NIJ_93} potentials we found close results and 
similar conclusions were also reached using separable interactions \cite{AF_PRC_89}.
To investigate the structure of this state, we looked at the two-body correlation functions.
The results (figure \ref{figcorrel}) show a superposition of two structures with different 
length scales, the short distance part being similar to the triton one.
This suggests -- rather than a breathing mode as it is widely considered in the literature --
a 3+1 structure formed by
an almost unperturbed triton plus a fourth nucleon orbiting around.
This state, with a r.m.s. $R\approx5$ fm, constitutes actually the first "halo-nucleus".
It seems difficult for a strong interaction to generate a first excitation in the continuum
and the open question is whether or not 
Coulomb forces, when added to the existing models,
would be able to put it in the right place with the required accuracy.

Within the same model we have calculated (table \ref{tab_lep}) the parameters of the effective range expansion 
\begin{equation}\label{mere}
g(k)\equiv k\;cot(\delta)=\left[{1-\left({k\over k_0}\right)^2}\right]^{-1}
\left[-{1\over a}+{1\over2}r_0k^2 +v_0k^4+o(k^6) \right]
\end{equation}
for N+NNN S-states.
Expansion (\ref{mere}) provides an accurate representation of the below breakup scattering although
the existence of the $^4$He$^*$ pole makes necessary the explicit inclusion of a singular term
for the T=S=0 channel. 
It is worth noticing
the coherence between the $^4$He$^*$ binding energy and the scattering parameters from table \ref{tab_lep}. 
If expansion (\ref{mere}) is inserted in the S-wave scattering amplitude $f_0^{-1}(k)=g(k)-ik$,
2 imaginary poles follow and
the nearest to threshold  $k_{-}=0.095$ fm$^{-1}$ corresponds to $E_-=-0.25$ MeV,
in agreement with the direct calculations of the $^4$He$^*$ binding. 

The 4N scattering states with two open channels N+NNN$\rightarrow$NN+NN
has been calculated for the S=T=0 final state \cite{CC_98}.
Of particular interest is the extraction of the imaginary part of the strong NN+NN scattering length
which controls the fusion rate in the process d+d~$\rightarrow$~n+$^3$He. We found
$a_R=+4.91\pm0.02$ fm and a very small value of $a_I=-0.0115\pm0.0001$ fm 
which should be only slightly modified once the Coulomb interaction is switched on.
This  value is due to the small overlapping between the K and H configurations which
respectively govern the N+NNN and NN+NN asymptotic states.
Other calculations of the dd cross section at very low energy
exist \cite{Oryu} but no values of $a_I$ were given.

\begin{figure}[hbtp]
\begin{minipage}[t]{64mm}
\epsfxsize=6.4cm\epsfysize=6.cm{\epsffile{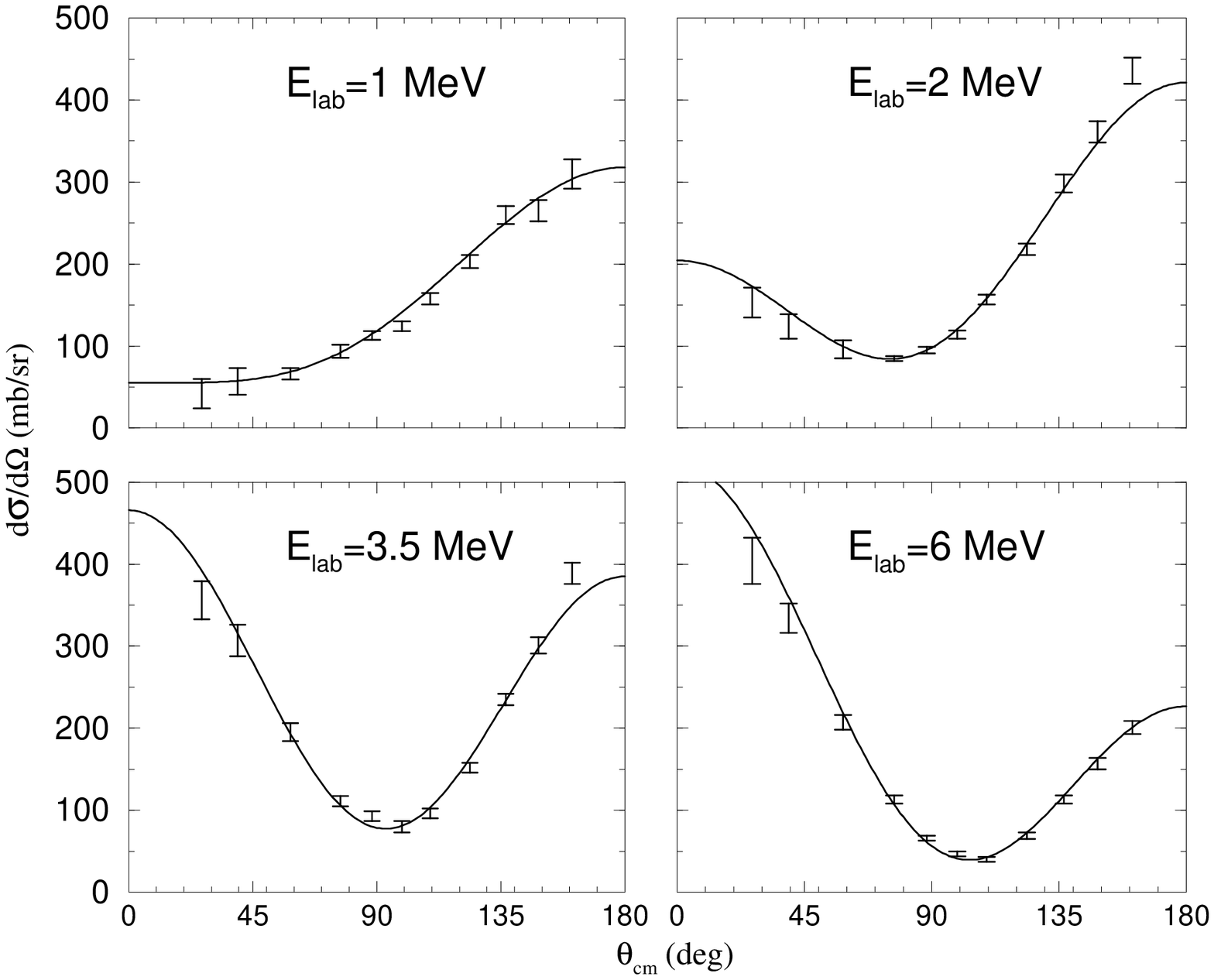}} 
\caption{n-t differential cross section with MT I-III at different n laboratory energies}\label{dcs_nt_MT13}
\end{minipage}
\hspace{0.1cm}
\begin{minipage}[t]{52mm}
\epsfxsize=5.2cm\epsfysize=6.0cm{\epsffile{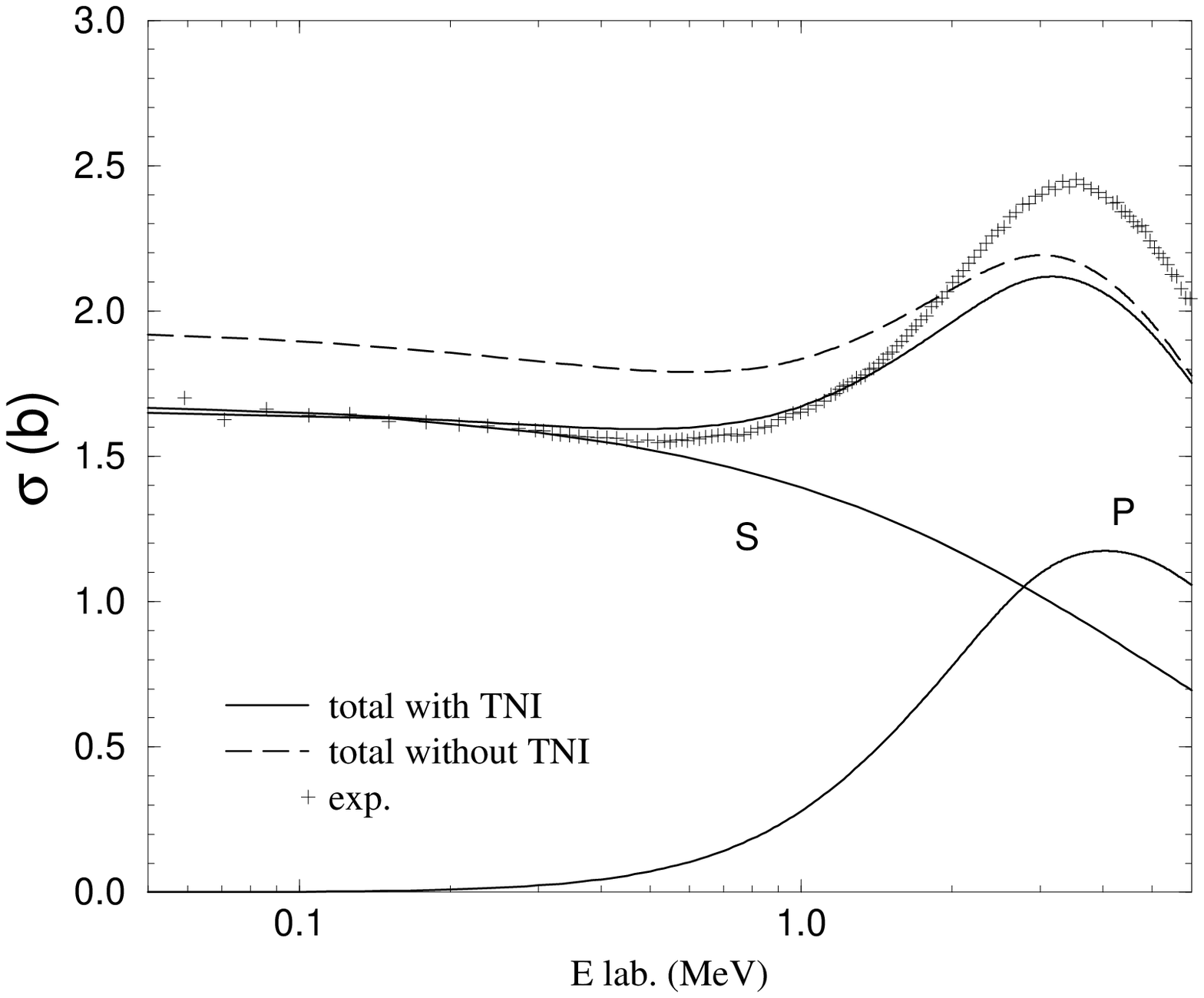}}
\caption{n-t total cross section with AV14 compared to experiment}\label{nt_av14}
\end{minipage}
\end{figure}

Special attention was paid to the n+t scattering. 
We found \cite{CC_98} that the description provided by the simple MT I-III model fits remarkably well
the data, as it can be seen in figure \ref{dcs_nt_MT13} where the differential cross sections 
at different energies are displayed.
Using realistic potentials, we have performed \cite{CCG_99} the same calculations 
for the $J^{\pi}=0^+,1^+,0^-,1^-,2^-$ states  
with $^1S_0,^3S_1,^3D_1$ waves in $V_{NN}$ and
expansion (\ref{KPW}) limited to $l_x,l_y,l_z=0,1,2$.
The singlet ($a_0$), triplet ($a_1$), coherent ($a_c$)
scattering length and zero energy cross section ($\sigma_0$) are in table \ref{tab_a_nt_realistic},
compared with experimental values taken from \cite{nt_exp}.
They are in very good agreement with those obtained by Pisa group \cite{Pisa}
$a_{0}=4.32$ and $a_{1}=3.80$ but the infered $a_c$ and $\sigma(0)$ values are well far from experiment. 
\begin{table}[htbp]
\begin{center}
\caption[]{n+t scattering length with realistic interactions}\label{tab_a_nt_realistic}
\begin{tabular}{lrrll}\hline
         & $a_{0}$   & $a_{1}$   & $a_c={1\over4}a_0+{3\over4}a_1$ & $\sigma(0)=\pi(a_0^2+3a_1^2)$ \\ \hline
AV14     &   4.31    & 3.79      &   3.92                          &   194                       \\  
Nijm II  &   4.31    & 3.76      &   3.90                          &   192 			    \\  
MT I-III &   4.10    & 3.63      &   3.75                          &   177  	                \\  
exp      &    --     & --        &   3.59$\pm$0.02                 &   170$\pm$ 3	                \\  
\end{tabular}
\end{center}
\end{table}

Calculations have been pursued beyond zero energy.
One can see from figure \ref{nt_av14} (dashed curve),
that two-body realistic potentials alone fail in describing the low energy cross section
as they failed in giving the binding energies $B_3$ and $B_4$. 
We would like to emphasize that this failure is in fact already present in the n+d case,
although somehow hidden
because the sizeable disagreement concerns only the doublet scattering length which turns
to be very small compared to the quartet one and it is furthermore suppressed by statistical factors.

To improve this situation we introduced the phenomenological TNI:
\[ W(\rho)=W_r {e^{-2\mu\rho}\over\rho}-W_a {e^{-\mu\rho}\over\rho} \qquad \rho=\sqrt{x^2+y^2} \]
Using the parameter set $ W_r=500 \quad W_a=174 \quad \mu=2.0$, we
got an overall low energy agreement with the values
$B_3$=8.48 MeV, $B_4$=29.0 MeV, $a_0$=4.0 fm, $a_1=3.53$ fm, $a_c=3.65$ fm and $\sigma(0)$=168 fm$^2$
and a satisfactory n+t S-wave cross section (solid line).
We point out that some dispersion exists in the experimental results which
makes difficult the comparison with theoretical values.
If one assumes the two last measurements \cite{nt_exp} of $a_c$ to be valid, 
our calculations are not far from the proposed value $a_c=3.60$. 
In this case, however, our scattering lengths are not compatible
with those infered from experiment, which are in their turn incompatible with each other.
A more precise measurement of $\sigma(0)$ would be helpful. For that purpose
the possibilities offered by the neutron beam facility, recently approved
at CERN \cite{CERN_TOF}, could deserve some attention from the nuclear Few-Body community.

If the inclusion of TNI is determinant in the n+t S-wave region, it doesn't improve the resonance peak,
on the contrary.
In a first tentative to explain this discrepancy we have included P-waves in the NN 
interaction \cite{CCG_99}.
We found that although separated P-waves had significant 
influence in the cross section, their global effect was very small and they 
could not explain the disagreement.
A recent work done by A. Fonseca \cite{F_99}
using AGS equations and 1-rank approximation, was able to push 
expansion (\ref{KPW}) further than we could
do and found a satisfactory description.
It would be interesting to have independent confirmation of this result and
 go beyond the first obstacle of the A=4 continuum.

\begin{acknowledge}
The numerical calculations have been performed with the Cray-T3E of the CGCV (CEA) and IDRIS (CNRS).
\end{acknowledge}

\end{document}